\documentstyle[12pt]{article}
\begin{document}
%%%%%%%%%%%%%%%%%%%%%%text%%%%%%%%%%%%%%%%%%%%%%
\def\thefootnote{\fnsymbol{footnote}}
\begin{flushright}
KANAZAWA-03-15  \\ 
July, 2003
\end{flushright}
%\vspace{ .7cm}
\vspace*{2cm}
\begin{center}
{\LARGE\bf  $\mu$ problem in an extended gauge mediation supersymmetry
 breaking}\\
\vspace{1 cm}
{\Large Daijiro Suematsu}
\footnote[1]{e-mail: suematsu@hep.s.kanazawa-u.ac.jp}
\vspace*{1cm}\\
{\it Institute for Theoretical Physics, Kanazawa University,\\
        Kanazawa 920-1192, Japan}\\    
\end{center}
\vspace{1cm}
{\Large\bf Abstract}\\  
%%%%%%%%%%%%%%%%%Abstract%%%%%%%%%%%%%%%%%%%%%%%%%%%
We study the $\mu$ problem and the radiative electroweak symmetry 
breaking in an extended gauge mediation supersymmetry breaking (GMSB) model,
in which the messenger fields are assumed to couple to the different
singlet fields due to the discrete symmetry. Since the spectrum of
superpartners is modified,
the constraint from the $\mu$ problem can be relaxed in comparison 
with the ordinary 
GMSB at least from the viewpoint of radiative symmetry breaking. 
We study the consistency of the values of $\mu$ and $B_\mu$ with 
the radiative electroweak symmetry breaking and also the mass spectrum 
of superpartners. 
We present such a concrete example in a supersymmetric SU(5) unified 
model which is constructed from a direct product gauge structure by 
imposing the doublet-triplet splitting. 
\newpage
\setcounter{footnote}{0}
\def\thefootnote{\arabic{footnote}}
%%%%%%%%%%%%%%%%%%%%%%text%%%%%%%%%%%%%%%%%%%%%%
\section{Introduction} 
Supersymmetry is now considered to be the most promising candidate
for the solution of the gauge hierarchy problem. Although we have no
direct evidence of the supersymmetry still now, the unification shown by the 
gauge couplings in the minimal supersymmetric standard model (MSSM)
may indirectly reveal its signal.
In the supersymmetric models the most important subject is to clarify
the supersymmetry breaking mechanism in the observable world.
Flavor changing neutral current processes severely constrain the
scenario for the supersymmetry breaking. From this point of view the
gauge mediation supersymmetry breaking (GMSB) 
\cite{mgm0}-\cite{mgmrev} seems to be prominent
since the mediation is performed in the flavor blind way by the standard 
model gauge interaction. 

In the ordinary minimal GMSB scenario \cite{mgm1}-\cite{exmgm2}, 
the messenger fields 
$(q, \ell)$ and $(\bar q, \bar\ell)$ which come from the vectorlike chiral 
superfields ${\bf 5}+\bar{\bf 5}$ of SU(5) are considered to have
couplings in the superpotential such as
\begin{equation}
W_{\rm GMSB}=\lambda_q Sq\bar q +\lambda_\ell S\ell\bar \ell,
\label{eqa}
\end{equation}
where $S$ is a singlet chiral superfield.\footnote{We will use the same
notation for the scalar component as its chiral superfield.} 
Both the scalar component $S$ 
and its $F$-term $F_S$ are assumed to get vacuum expectation 
values (VEVs) through the couplings to the fields in the hidden 
sector where supersymmetry is assumed to be broken. 
The masses of the gauginos and the scalar superpartners are respectively 
produced by the one-loop and two-loop effects through the
couplings in eq.~(\ref{eqa}). These masses are characterized by 
$\Lambda\equiv \langle F_S\rangle/\langle S\rangle$ and then 
$\Lambda$ is considered to be in the range 20~-~100~TeV.

The chiral superfield $S$ is usually considered not to have a direct 
coupling to the doublet Higgs chiral superfields $H_1$ and $H_2$ 
in the superpotential, although it can be an origin for both the $\mu$
term and the bilinear soft supersymmetry breaking parameter $B_\mu$. 
The reason is that the relation $B_\mu=\Lambda\mu$ induced from 
such a coupling makes $\vert B_\mu\vert$ too large for the electroweak symmetry
breaking under the assumption $\mu=O(100)$~GeV. On the other hand, 
if we assume that $\vert B_\mu\vert$ has a suitable value for the
electroweak symmetry breaking, the resulting small $\mu$ cannot satisfy
the potential minimum condition.\footnote{In the ordinary GMSB scenario 
the potential minimum condition
requires $\vert B_\mu\vert <\mu^2$ as seen later.} 
Since $\Lambda$ takes a large value as mentioned above, it makes 
both $\mu$ and $B_\mu$ difficult to take suitable values for the
radiative symmetry breaking \cite{mgm1,mgm1a,mgm2,mgmrev}.
Even if there are no such a coupling in the superpotential, 
$\mu$ and $B_\mu$ can be produced radiatively picking up the
supersymmetry breaking effect and 
$B_\mu=\Lambda\mu$ is generally satisfied \cite{muproblem}. 
This suggests that the electroweak symmetry breaking cannot be 
induced radiatively also in this case.
Thus, it is usually considered that the $\mu$ term should
have another independent origin. This requires us to 
introduce new additional fields for this purpose. A lot of 
models of this kind have been proposed by now 
\cite{mgm1,mgm1a,mgm2,mgmrev,muproblem}.

In this paper we show that an extended GMSB model proposed here may make ease
the difficulty of the $\mu$ problem in comparison with the ordinary 
GMSB at least from the viewpoint of the radiative symmetry
breaking. In this model we can relax the constraint on $\mu$ and
$B_\mu$.  
The consistency of such a scenario with the radiative electroweak 
symmetry breaking is studied in some detail by solving numerically the
renormalization group equations (RGEs). Its phenomenological results are 
also discussed. We present its concrete example in the deconstructed SU(5) 
unified model. In this model the superpotential is suitably arranged 
by the discrete symmetry which is introduced to resolve the 
doublet-triplet Higgs degeneracy \cite{dt} in the basis of the direct product 
gauge structure. 

\section{Soft SUSY breaking and $\mu$ problem}

We extend the superpotential (\ref{eqa}) for the messenger fields  
in such a way that the messenger fields 
$q, \bar q$ and $\ell, \bar\ell$ couple to
the different singlet chiral superfields $S_1$ and $S_2$ which are
assumed to have couplings to the hidden sector where the supersymmetry 
is supposed to be broken. This can happen incidentally as a result of 
a suitable discrete 
symmetry as we will see it later in an explicit example.
Thus the couplings of messenger fields are expressed as
\begin{equation}
W_{\rm GMSB}^\prime= \lambda_q S_1q\bar q + \lambda_\ell S_2\ell\bar\ell.
\label{eqdd}
\end{equation}  
If we assume that both $S_\alpha$ and $F_{S_\alpha}$ 
get the VEVs, the gaugino masses and the soft scalar masses are 
generated through the one-loop and two-loop diagrams, respectively, as in
the ordinary case. 
However, the mass formulas are modified from the ordinary ones
since the messenger fields $q,~\bar q$ and $\ell,~\bar\ell$ 
couple to the different singlet fields $S_\alpha$.

The mass formulas of the superpartners in this type of model 
have been discussed in \cite{sue}.
Under the ordinary assumption such as
$\langle F_{S_\alpha}\rangle\ll \lambda_{q,\ell}\langle
S_\alpha\rangle^2$ \cite{mgm1}, the mass formulas take a very simple form.
The masses $M_r$ of the gauginos $\lambda_r$ of the MSSM gauge group 
can be written in the form as 
\begin{equation}
M_3={\alpha_3\over 4\pi}\Lambda_1, \qquad
M_2={\alpha_2\over 4\pi}\Lambda_2, \qquad
M_1={\alpha_1\over 4\pi}\left({2\over 3}\Lambda_1+\Lambda_2\right), 
\label{eqff}
\end{equation}
where $\alpha_r=g_r^2/4\pi$ and 
$\Lambda_\alpha=\langle F_{S_\alpha}\rangle/\langle S_\alpha\rangle$.
The soft scalar masses $\tilde m_f^2$ can be written as
\begin{equation}
\tilde m^2_f=2\left[C_3\left({\alpha_3\over 4\pi}\right)^2 
+{2\over 3}\left({Y\over 2}\right)^2\left({\alpha_1\over 4\pi}\right)^2\right]
\vert\Lambda_1\vert^2 
+2\left[C_2\left({\alpha_2\over 4\pi}\right)^2 
+\left({Y\over 2}\right)^2\left({\alpha_1\over 4\pi}\right)^2\right]
\vert\Lambda_2\vert^2,
\label{eqg}
\end{equation}
where $C_3=4/3$ and 0 for the SU(3) triplet and singlet fields, and
$C_2=3/4$ and 0 for the SU(2) doublet and singlet fields, respectively. 
The hypercharge $Y$ is expressed as $Y=2(Q-T_3)$.
 
These formulas can give a rather different mass spectrum for the gauginos
and the scalar superpartners in comparison with the ordinary GMSB scenario.
The spectrum depends on the value of $\Lambda_2/\Lambda_1$. 
In fact, if we assume $\Lambda_1< \Lambda_2$, the mass difference
between the color singlet fields and the colored fields tends to be 
smaller in comparison with the one in the ordinary scenario at least in
the supersymmetry breaking scale.
As an example, we take $\Lambda_1=60$~TeV and $\Lambda_2=150$~TeV to
show a typical spectrum of the superpartners at the supersymmetry
breaking scale. The resulting spectrum is
\begin{eqnarray}
&& M_3\simeq 415~{\rm GeV}, \quad   
M_2\simeq 418~{\rm GeV}, \quad 
M_1\simeq 166~{\rm GeV}, \quad
\tilde m_Q\simeq 851~{\rm GeV}, \nonumber \\
&& \tilde m_U\simeq 690~{\rm GeV}, \quad 
\tilde m_D\simeq 682~{\rm GeV}, \quad  
\tilde m_L\simeq 520~{\rm GeV}, \quad 
\tilde m_E\simeq 195~{\rm GeV}, \nonumber \\ 
&&m_1=m_2\simeq 520~{\rm GeV},  
\label{eqi}
\end{eqnarray}
where $m_1$ and $m_2$ are masses of the Higgs scalars that couple 
with the fields in the down and up sectors of quarks and leptons, respectively.
These masses are somewhat affected by the running effect based on the
renormalization group equations (RGEs), although the running region 
depends on the values of $\Lambda_1$ and $\Lambda_2$. 
As discussed in \cite{exmgm2}, in the minimal GMSB model 
the soft supersymmetry breaking $A_f$ parameters can also be expected
to be induced through the radiative correction in such a way as
\begin{equation}
A_f\simeq A_f(\Lambda)+M_2(\Lambda)\left(-1.85+0.34\vert h_t\vert^2\right)
+\cdots,
\label{eqj}
\end{equation}
where we should omit a term of $h_t$ except for the top sector $(f=t)$. 
Thus, even if $A_f(\Lambda)=0$ is assumed, we can expect $A_f$ to be
generated through this effect.\footnote{
In the present study we assume $A_f(\Lambda)=0$.
The soft supersymmetry breaking
parameters $B_\mu/\mu$ is also known to follow the similar radiative
correction to eq.~(\ref{eqj}) and the phenomenological studies have been
done \cite{exmgm2, b0, b0a}. However,
we will discuss the origin of $B_\mu(\Lambda)$ in the followings.}

As mentioned in the introduction, the values of $\mu$ and $B_\mu$ 
are crucial for the electroweak symmetry breaking.
Here we examine the effect of the introduction of a 
coupling $\lambda_\mu S_1H_1H_2$ in the superpotential. 
It can give a contribution to both $\mu$ and $B_\mu$ terms in the 
form as
\begin{equation}
\mu=\lambda_\mu\langle S_1\rangle, \qquad
B_\mu=\lambda_\mu\langle F_{S_1}\rangle.
\label{eqd}
\end{equation} 
Unfortunately, as in the ordinary case the problematic relation 
$B_\mu=\mu\Lambda_1$ is satisfied also in this case. 
However, this relation does not eventually rule out the possibility for
the radiative symmetry breaking in the present case. This is very
different from the ordinary GMSB.

An important aspect of the $\mu$ problem in the GMSB is crucially 
related to the radiative electroweak symmetry breaking. 
In order to see this, we study the well-known conditions for the 
radiative electroweak symmetry breaking.
In the MSSM the minimization conditions of the tree-level scalar
potential are written as
\begin{eqnarray}
&&\sin 2\beta={2 B_\mu \over m_1^2 +m_2^2+ 2\mu^2},  
\label{eqe0} \\
&&m_Z^2={2m_1^2-2m_2^2\tan^2\beta\over \tan^2\beta-1}-2\mu^2,
\label{eqe}
\end{eqnarray}
where we assume that $\mu$ and $B_\mu$ are real, for simplicity.
In these equations the Higgs scalar masses $m_1^2$ and $m_2^2$ should 
be improved into the values at the weak scale by using the RGEs. 
If we take account of the dominant one-loop
contributions, they can be written as \cite{exmgm2}
\begin{eqnarray}
&&m_1^2(M_W)\simeq m_1^2(\Lambda)-{3\over 2}M_2^2(\Lambda)
\left({\alpha_2(M_W)^2\over \alpha_2(\Lambda)^2}-1\right) 
-{1\over 22}M_1^2(\Lambda)
\left({\alpha_1(M_W)^2\over \alpha_1(\Lambda)^2}-1\right), \nonumber \\
&&m_1^2(M_W)-m_2^2(M_W)\simeq {6 h_t^2\over 8\pi^2}m_{\tilde t}^2
\ln\left(\Lambda\over m_{\tilde t}\right),
\label{eqf}
\end{eqnarray}
where $h_t$ and $m_{\tilde t}$ represent the top Yukawa coupling
constant and the stop mass. They are approximated by the values at $\Lambda$.
The masses of the gauginos and scalar superpartners at the 
supersymmetry breaking scale $\Lambda$ are determined by
eqs.~(\ref{eqff}) and (\ref{eqg}).

We first remind the situation for the radiative symmetry breaking 
in the ordinary GMSB case ($\Lambda_1=\Lambda_2$) by checking a condition
$m_1^2+m_2^2+2\mu^2>2\vert B_\mu\vert$, which is obtained from the condition 
(\ref{eqe0}) and is also required by the vacuum stability. 
Inserting eq.(\ref{eqf}) into this inequality, we find 
that this necessary condition can be approximately written as
\begin{equation}
\left[{3\over 2}
\left(\alpha_2\over 4\pi\right)^2
+{5\over 6}\left(\alpha_1\over 4\pi\right)^2\right]
\left(2-{4h^2_t\over 3\pi^2}\left({\alpha_3\over\alpha_2}\right)^2
\ln{\sqrt 6\pi\over \alpha_3}\right) 
>{2\over\Lambda^2}(\vert B_\mu\vert -\mu^2). 
\label{eqh}
\end{equation}
It is easy to find that the condition (\ref{eqh}) is never satisfied 
unless $h_t$ takes an unacceptably small value in the case of 
$\vert B_\mu\vert>\mu^2$, which is caused by
the relation $B_\mu=\mu\Lambda$ because of $\Lambda\gg \mu$. 
Thus, we need to consider an additional origin of $\mu$ 
to make the condition $B_\mu<\mu^2$ be satisfied.
This is the well-known result in the ordinary GMSB scenario 
\cite{mgm2,mgmrev}. 
This fact might make us consider that the condition
$m_1^2+m_2^2+2\mu^2>2\vert B_\mu\vert$ cannot be satisfied in the GMSB model
without the new origin for $\mu$ as far as the undesirable relation 
$B_\mu=\mu\Lambda$ exists.
In the present model, however, $\Lambda_1$ is not generally supposed 
to be equal to $\Lambda_2$. 
This feature can give us a new possibility with regard to 
the electroweak symmetry breaking even under the existence of 
the relation $B_\mu=\mu\Lambda_1$, if $\Lambda_1 < \Lambda_2$ 
is satisfied and then the spectrum of superpartners is modified.

In order to see this, it is useful to note that
the factor $(\alpha_3/\alpha_2)^2\ln(\sqrt 6\pi/\alpha_3)$ 
in eq.~(\ref{eqh}) should be modified into an approximated factor
\begin{equation}
\left[1+{16\over 9}\left(\alpha_3\over\alpha_2\right)^2
\left(\Lambda_1\over\Lambda_2\right)^2\right]
\ln{\sqrt 6\pi\over (\alpha_3^2(\Lambda_1/\Lambda_2)^2+9\alpha_2^2/16)^{1/2}}
\end{equation}
in the present extended GMSB. 
This is caused by the change in the formulas of the soft scalar masses.
We find that $m_1^2+m_2^2+2\mu^2>2\vert B_\mu\vert$ is
satisfied as far as $\Lambda_1<\Lambda_2$ 
even in the case of $h_t\simeq 1$ and 
$\vert B_\mu\vert >\mu^2$.\footnote{In this paper we assume that
$\Lambda_{1,2}$ and $\mu$ are positive.}
The same change related to the radiative correction due to the top
Yukawa coupling tends to make the allowed value of $\tan\beta$ smaller 
than the one in the ordinary GMSB scenario with the additional $\mu$
contribution \cite{exmgm2,b0a,mgm2}. This can be found from eq.~(\ref{eqe}).   
Moreover, the same equation suggests that there appears an upper bound 
of $\Lambda_2/\Lambda_1$ if we impose the lower bound for $\tan\beta$. 
For example, if we require $\tan\beta>2$, 
we find that $\Lambda_2/\Lambda_1~{^<_\sim}~3.5$ should be satisfied.  

The more accurate analysis on this aspect can be done numerically 
by using the one-loop RGEs in the MSSM. For this purpose we can transform the 
conditions (\ref{eqe0}) 
and (\ref{eqe}) into the formulas for $\mu^2$ such as
\begin{eqnarray}
&&\mu^2_A={1\over 4\Lambda_1^2}
\left[(m_1^2-m_2^2)\tan 2\beta+m_Z^2\sin 2\beta\right]^2, \nonumber \\
&&\mu^2_B=(m_1^2-m_2^2){\tan^2\beta\over \tan^2\beta -1}-m_1^2-{1\over 2}m_Z^2,
\end{eqnarray}
where we use $B_\mu=\mu\Lambda_1$. 
To estimate these formulas we take the following procedure. 
The gauge and Yukawa coupling constants are
evolved from the gauge coupling unification scale to the weak scale.
The soft supersymmetry breaking parameters are introduced at $\Lambda_2$
and evolved to the weak scale. 
We use the weak scale values of $m_1^2$ and $m_2^2$ obtained in this way
and also the value of $\tan\beta$ which is determined by the top quark 
mass and the value of top Yukawa coupling obtained from the RGEs.  
In Fig.~1 we plot each value of $\mu^2_A$ and $\mu_B^2$ in the 
$(\Lambda_2/\Lambda_1,~\mu^2)$ plane for $\Lambda_1=60$~TeV and several
values of $\tan\beta$. $\mu^2_A$ takes very small values of $O(1)$~GeV
and the smaller $\tan\beta$ realizes the larger value of $\mu_A^2$.
$\mu_B^2$ is very sensitive to the value of $\Lambda_2/\Lambda_1$ in
comparison with $\mu_A^2$.
From this figure we can find that there are solutions in the region such as 
$\Lambda_2/\Lambda_1~{^<_\sim}~3$ if we impose $\tan\beta~{^>_\sim}~2.4$ 
which corresponds to the constraint from the neutral Higgs boson search.
However, if we take the larger value for $\Lambda_1$, we can obtain 
the solutions for the larger values of $\Lambda_2/\Lambda_1$.
%%%%
\input epsf 
\begin{figure}[tb]
\begin{center}
\epsfxsize=8cm
\leavevmode
\epsfbox{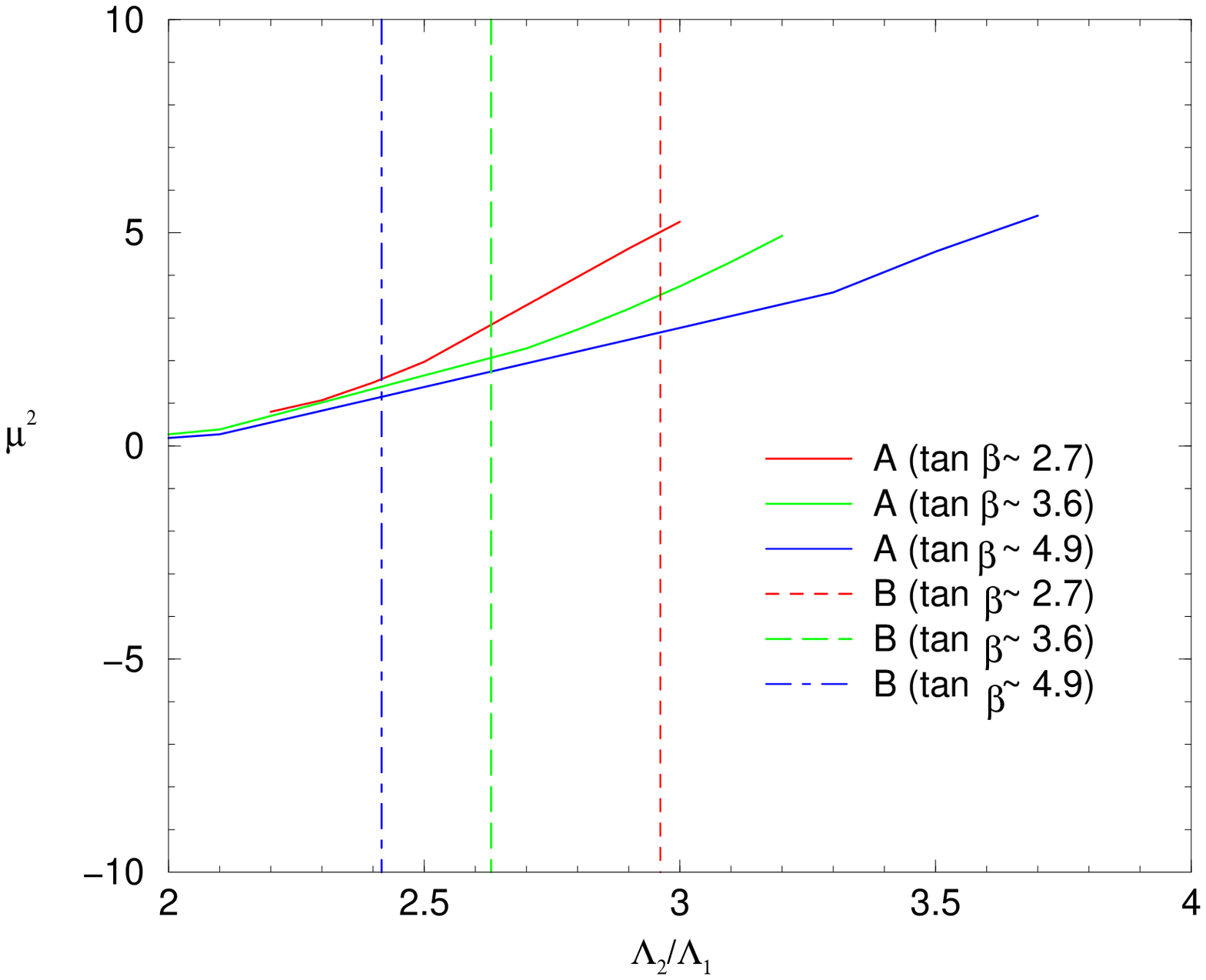}
\vspace*{-0.3cm}
\end{center}
{\footnotesize Fig. 1~~\  The values of $\mu^2_A$ and $\mu_B^2$ 
predicted by the conditions for the radiative symmetry breaking in the 
case of $B_\mu=\mu\Lambda_1$. We can find the solutions as the crossing
 points of $\mu_A^2$ and $\mu_B^2$.}
\end{figure}
%%%%%

Although the radiative symmetry breaking can be found to occur just within the
framework without adding any fields, we need to impose other
phenomenological constraints for the scenario to be realistic.
Since the absolute values of $\Lambda_1$ and $\mu$ are directly constrained 
by the experimental bounds for the masses of the gluino and 
the neutralino, 
we find that these values should be in the range
\begin{equation}
\Lambda_1~{^>_\sim}~20~{\rm TeV}, \qquad \mu
~{^>_\sim}~ 10^{2}~{\rm GeV}.
\end{equation} 
This means that the present solution to the $\mu$ problem requires other
contribution to the $\mu$ term to overcome the constraint from the 
neutralino mass bound. However, it is useful to note that 
the situation on the origin of $\mu$ term is not the same as the 
ordinary case. Since $\vert B_\mu\vert <\mu^2$ is not required in
this case unlike the ordinary GMSB, the constraint imposed 
from the radiative symmetry breaking 
on the additional contribution $\mu^\prime$ to the $\mu$ term 
can be expected to be sufficiently weaken.

In order to study this aspect, we study the radiative symmetry 
breaking and the spectrum of the 
superpartners by using the one-loop RGEs. In this study we need only 
modify the $\mu$ parameter into $\tilde\mu=\mu+\mu^\prime$, where $\mu$
and $B_\mu$ are defined by eq.~(\ref{eqd}). 
Since in the present soft supersymmetry breaking scheme there are four 
parameters and we take them as $\mu$ and $\mu^\prime$ in addition 
to $\Lambda_{1,2}$, we can predict the spectrum of the superpartners 
in a rather restrictive way through this study. 
As the phenomenological constraints, we impose the experimental mass
bounds for the superpartners and also require both the color and
electromagnetic charge not to be broken. Under these conditions we
search the allowed parameter region in the case of $\Lambda_1=60$~TeV. 
In Fig.~2 we give scatter plots for each value of $\tilde\mu$ and $B\equiv
B_\mu/\tilde\mu$ for the solutions of the radiative symmetry breaking 
at each value of $\Lambda_2/\Lambda_1$. 
In this figure we can see that there are the solutions 
with $\vert B_\mu\vert >\tilde\mu^2$ for the
$\Lambda_2/\Lambda_1~{^>_\sim}~2$ region, although the solutions are
restricted into the ones with $\vert B_\mu\vert <\tilde\mu^2$
for $\Lambda_2/\Lambda_1~{^<_\sim}~ 2$. 
Here the ordinary GMSB should be noted to
correspond to $\Lambda_2/\Lambda_1=1$. It should be also noted that 
the $\mu$ parameter is allowed to be smaller than the one in the 
ordinary GMSB.

We give the spectrum of the superpartners obtained in the same analysis 
for the case of $\Lambda_1=60$~TeV in Fig.~3.
On the lightest chargino and neutralino by combining Figs.~2 and 3
we can find that they are dominated by the gaugino in the region
$\Lambda_2/\Lambda_1~{^<_\sim}~2$ and they change into the 
Higgsino dominated one 
in the region $\Lambda_2/\Lambda_1~{^>_\sim}~2$.
The next lightest superpartner is always the neutralino as far as
$\Lambda_2/\Lambda_1>1$ is assumed. 
The CP-even neutral Higgs boson mass slightly decreases when 
$\Lambda_2/\Lambda_1$ increases. This follows the behavior of the stop mass. 
Although the neutral Higgs boson mass is almost equal 
to the experimental bound for this $\Lambda_1$ value,
it can be larger by taking $\Lambda_1$ larger.
The difference of the mass spectrum of superpartners from the one in the
ordinary GMSB becomes clear in the larger $\Lambda_2/\Lambda_1$ region.
In that region the mass gap between the colored fields and the 
color singlet fields becomes smaller and by using this feature 
we might distinguish the
present model from the ordinary one. 
%%%%
\begin{figure}[tb]
\begin{center}
\epsfxsize=8cm
\leavevmode
\epsfbox{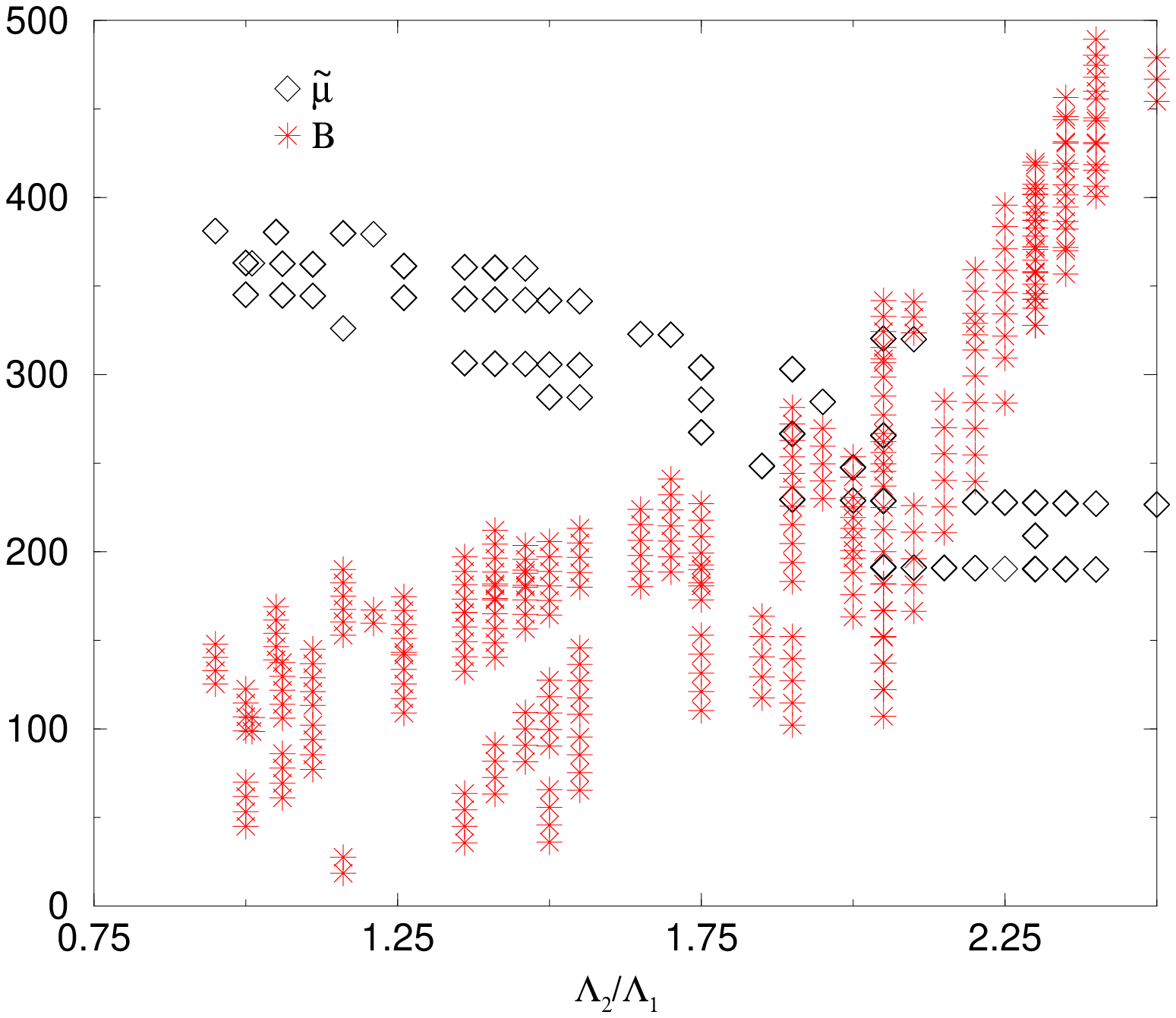}
\vspace*{-0.3cm}
\end{center}
{\footnotesize Fig. 2~~\  The relation 
between $\tilde\mu$ and $B\equiv B_\mu/\tilde\mu$} in the solutions for 
the radiative symmetry breaking conditions. The $\tilde\mu$ and $B$ is
 represented by the GeV unit.
\end{figure}
%%%%
\begin{figure}[tb]
\begin{center}
\epsfxsize=7cm
\leavevmode
\epsfbox{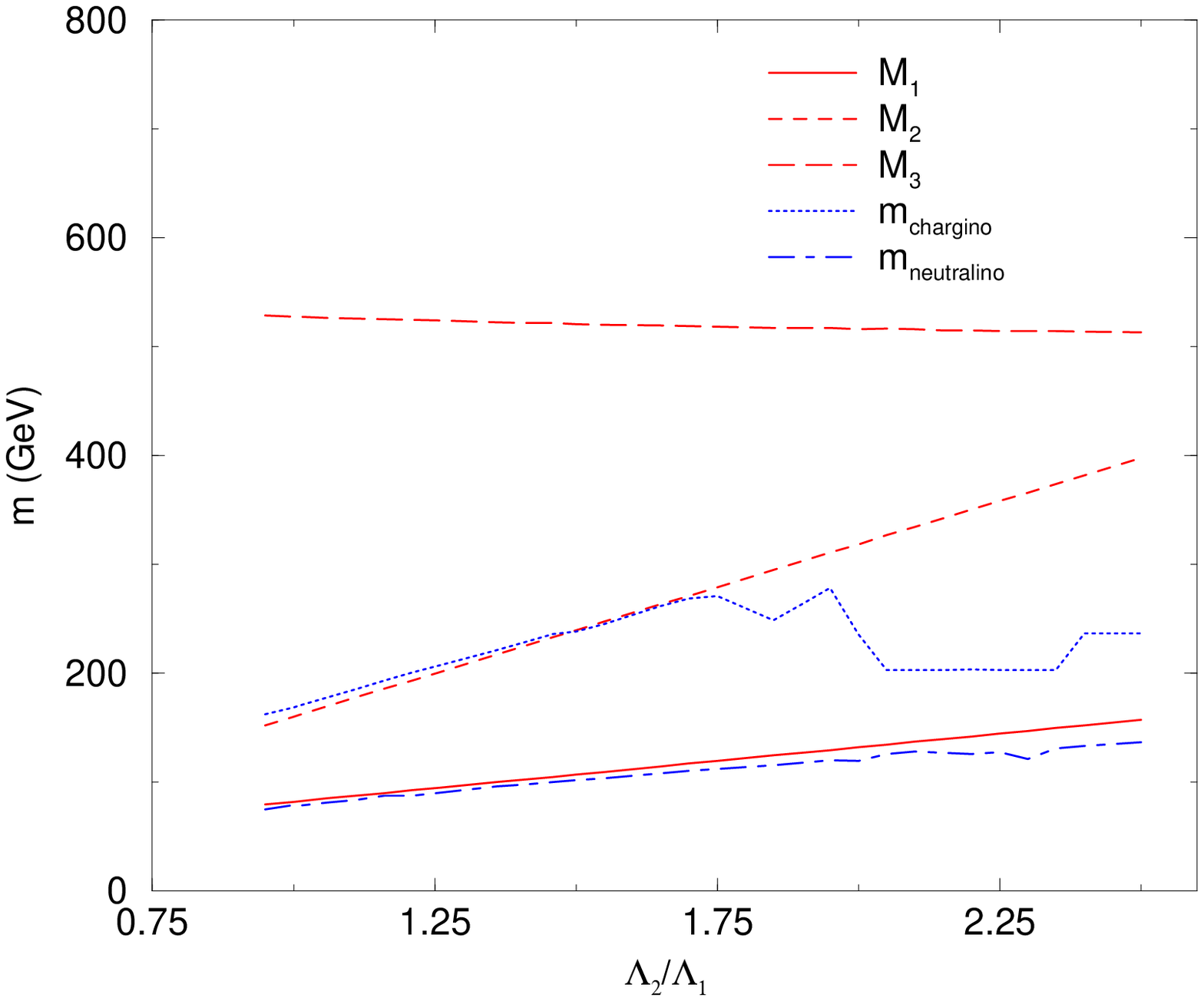}
\hspace*{7mm}
\epsfxsize=7cm
\leavevmode
\epsfbox{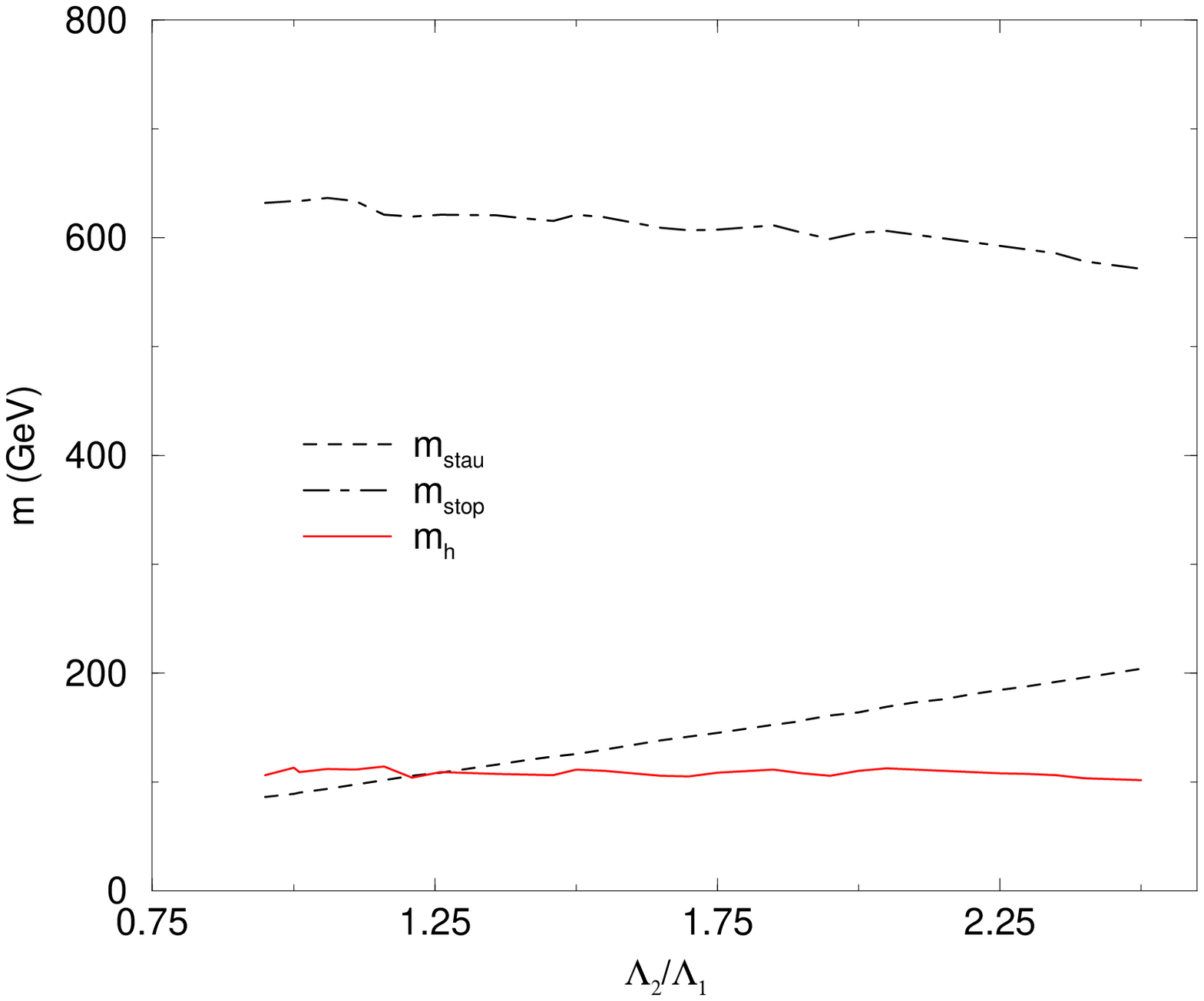}
\end{center}
\vspace*{-0.3cm}
{\footnotesize Fig. 3~~\  Mass spectrum of the superpartners for the
parameter sets which satisfy the radiative symmetry breaking conditions
and various phenomenological constraints. The lightest one for each
 superpartner is shown
 except for the gauginos.}
\end{figure}
%%%%%

Finally we briefly comment on other features which are not mentioned before.
Since the soft masses of two Higgs fields $H_1$ and $H_2$ are 
same at the supersymmetry breaking scale, the radiative symmetry
breaking predicts the relatively small value of $\tan\beta$ such as 
2.5 -- 7.5. Although $\tan\beta>10$ is possible, it needs the fine
tuning of parameters. 
There can appear an interesting feature for the coupling unification
scale in the case of $\Lambda_2/\Lambda_1 > 1$.
The unification scale of the coupling constants of SU(3) 
and SU(2) can be pushed 
up to the higher scale depending on the values of 
$\Lambda_{1,2}$.\footnote{The similar possibility has been discussed in 
other context in \cite{unif}.}  
This aspect comes from the fact that the SU(2) nonsinglet superpartners 
decouple earlier than others. However, we need further study whether the large
shift of unification scale can be consistent with the radiative symmetry 
breaking.

\section{A deconstructed SUSY SU(5) model}

In this section we consider a model which can realize the extended 
GMSB discussed 
in the previous section. As such an interesting example, we propose
a unified SU(5) model with a direct product gauge structure such as 
${\cal G}=$SU(5)$^\prime\times$SU(5)$^{\prime\prime}$ and a
global discrete symmetry $F$ which commutes with this gauge symmetry 
\cite{sue}.
A field content of the model is listed in Table 1.
They are composed of bifundamental chiral superfields 
$\Phi_1(\bar{\bf 5},{\bf 5})$ and $\Phi_2 ({\bf 5}, \bar{\bf 5})$,
an adjoint Higgs chiral superfield $\Sigma({\bf 1}, {\bf 24})$, 
three sets of chiral superfields 
$\Psi_{\bf 10}({\bf 10}, {\bf 1})+\Psi_{\bar{\bf 5}}(\bar{\bf 5}, {\bf 1})$ 
which correspond to three generations of
quarks and leptons, a set of chiral superfield 
$H({\bf 5}, {\bf 1})+\tilde H({\bf 1}, \bar{\bf 5})$ which contains 
Higgs doublets, and also a set of chiral superfield 
$\bar\chi(\bar{\bf 5}, {\bf 1})+\chi({\bf 1}, {\bf 5})$
in order to cancel the gauge anomaly induced by the above contents. 
We additionally introduce several singlet chiral superfields
$S_\alpha$ and $N$.

In order to induce the symmetry breaking at the high energy scale 
we introduce a superpotential such as
\begin{equation}
W_1= M_\phi{\rm Tr}\left(\Phi_1\Phi_2\right) 
+{1\over 2}M_\sigma{\rm Tr}\left(\Sigma^2\right)
+\lambda~{\rm Tr}\left(\Phi_1\Sigma\Phi_2 +{1\over 3}\Sigma^3\right). 
\end{equation}
As it is shown in \cite{sue}, it is easily found that the scalar
potential derived from this $W_1$ 
has a nontrivial minimum which is realized at
\begin{equation}
\sigma=\tilde M~{\rm diag}~(2,~2,~2,~-3,~-3), \qquad
\phi_1=\kappa\sigma, \qquad 
\phi_2={1\over \kappa}\left({M_\sigma\over M_\phi}-1\right)\sigma.
\label{eqb}
\end{equation}
where $\tilde M$ is defined as $\tilde M=M_\phi/\lambda$.
There is no $D$-term contribution to $V$ from these vacuum expectation
values (VEVs) in eq.~(\ref{eqb}) and then the
supersymmetry is conserved at this stage. 
The remaining symmetries and a parameter $\kappa$ can be determined 
by assuming that the model is constructed based on the suitable 
deconstruction \cite{sue}.

We consider that the theory space of the model is represented 
by the moose diagram which is composed of the $n$ sites $Q_i$ placed 
on the vertices of an $n$-polygon and one site on its center $P$ of 
this polygon \cite{w}.
We assign SU(5)$^\prime$ on the site $P$ and SU(5)$^{\prime\prime}$ on 
each site $Q_i$ and also put a bifundamental chiral superfield $\Phi_i$
on each link from $P$ to $Q_i$. On each link from $Q_i$ to $Q_{i+1}$ 
we put the adjoint Higgs chiral superfield $\Sigma$ of SU(5)$^{\prime\prime}$. 
Here we introduce an equivalence relation only for the boundary points of 
the polygon by the $2\pi/n$ rotation and we identify this $Z_n$ symmetry 
with the above mentioned discrete symmetry $F$.
This makes us consider the reduced theory space composed of only three
sites $P$, $Q_1$ and $Q_2$, in which the field contents become equivalent 
to the one given in Table 1.
Under these settings we can find the following interesting results 
\cite{w,sue}.
First, in this model the symmetry ${\cal G}\times F$ 
breaks down into ${\cal H}\times F^\prime$ by considering the vacuum 
defined by eq.~(\ref{eqb}).
Here ${\cal H}$=SU(3)$\times$SU(2)$\times$U(1) is a
subgroup of the diagonal sum SU(5) of ${\cal G}$ and also
a discrete symmetry $F^\prime$ is a diagonal subgroup of 
$F\times G_{{\rm U(1)}^{\prime\prime}}$ where
$G_{{\rm U(1)}^{\prime\prime}}$ is a discrete subgroup of a hypercharge
in SU(5)$^{\prime\prime}$.
Since the definition of $F^\prime$ contains the discrete subgroup of 
U(1)$^{\prime\prime}$ in SU(5)$^{\prime\prime}$ as its component,
every field which has a nontrivial transformation property with respect 
to SU(5)$^{\prime\prime}$ can have different charges. 
This feature makes it possible to split the doublet Higgs fields from
the colored Higgs fields and also forbid the messenger fields to couple
with the same singlet chiral superfields.
We assign the charges of $F^\prime$ for every field as shown in Table 1.
The second result is that the parameter $\kappa$ is fixed through an equation
$\kappa^2-\kappa+1-M_\sigma/M_\phi=0$. 
\begin{figure}[tb]
\begin{center}
\begin{tabular}{|l|c|c|c|c|}\hline
         & ${\cal F}({\cal G}~{\rm rep.})$  &  F    
& \multicolumn{2}{|c|}{$F^\prime$} \\\cline{4-5}
   &&& ${\bf 3}\in {\bf 5}$ or $\bar{\bf 3}\in\bar{\bf 5}$ & 
${\bf 2}\in{\bf 5}$ or $\bar{\bf 2}\in\bar{\bf 5}$ \\\hline
{\rm Quarks/Leptons}& $\Psi_{\bf 10}^j({\bf 10}, {\bf 1})$ &  $\alpha$ & 
$\alpha$ & $\alpha$ \\
$(j=1\sim 3)$   & $\Psi_{\bar{\bf 5}}^j(\bar{\bf 5}, {\bf 1})$ 
&$\beta$ & $\beta$ & $\beta$ \\
 & $\Psi_{\bf 1}^j({\bf 1}, {\bf 1})$ &$\gamma$ &$\gamma$ & $\gamma$  \\\hline 
{\rm Higgs fields}& $H({\bf 5}, {\bf 1})$ & $\rho$ & $\rho$ & $\rho$\\
           & $\tilde H({\bf 1}, \bar{\bf 5})$ & $\xi$ 
& $\xi+2a$ & $\xi-3a$ \\ \hline
{\rm Messenger fields}& $\bar\chi(\bar{\bf 5}, {\bf 1})$ & $\delta$ & 
$\delta$ & $\delta$\\ 
              & $\chi({\bf 1}, {\bf 5})$ & $\epsilon$ & 
$\epsilon+2b$ & $\epsilon-3b$
 \\ \hline
{\rm Bifundamental field}  & $\Phi_1(\bar{\bf 5},{\bf 5})$ & $\eta$ 
& $\eta+2c$ &  
$\eta-3c$ \\
            & $\Phi_2({\bf 5},\bar{\bf 5})$ & $\zeta$ & $\zeta+2d$ & 
$\zeta-3d$ \\\hline
{\rm Adjoint Higgs field}  & $\Sigma({\bf 1},{\bf 24})$ & 0 & 
\multicolumn{2}{|c|}{0}     \\\hline
{\rm Singlets}   & $S_1({\bf 1}, {\bf 1})$& $\theta$ &
 \multicolumn{2}{|c|} {$\theta$}\\
                 &$S_2({\bf 1}, {\bf 1})$ & $\tau$ & 
\multicolumn{2}{|c|}{$\tau$}\\
                 &$N({\bf 1}, {\bf 1})$ & $\omega$ & 
\multicolumn{2}{|c|}{$\omega$}\\ \hline
\end{tabular}
\vspace*{3mm}\\
{\footnotesize Table 1~ Charge assignment of the discrete symmetry 
$F^\prime$ for the chiral superfields.} 
\end{center}
\end{figure} 

In order to fix the discrete symmetry $F^{\prime}$ we impose the 
following conditions on $F^\prime$ to satisfy various 
phenomenological constraints, as we have done it in \cite{sue}.\\
(i)~Each term in the superpotential $W_1$ should exist and this
requirement imposes the conditions:
\begin{equation}
\eta+\zeta +2(c+d)=0, \qquad \eta+\zeta -3(c+d)=0.
\label{con1}
\end{equation}
(ii)~To realize the doublet-triplet splitting, only the color triplet 
Higgs chiral superfields $H_{\bf 3}$ and $\tilde H_{\bf 3}$ except for the
ordinary doublet Higgs chiral superfields $H_{\bf 2}$ and $\tilde H_{\bf 
2}$ should become 
massive. Thus the Yukawa coupling $\Phi_1 H_{\bf 2}\tilde H_{\bf 2}$ should be 
forbidden although $\Phi_1H_{\bf 3}\tilde H_{\bf 3}$ is allowed. 
This gives the conditions such as
\begin{equation}
\rho+\xi+\eta-3(a+c)\not=0,\qquad \rho+\xi+\eta+2(a+c)= 0.
\end{equation}
(iii)~Yukawa couplings of quarks and leptons, that is, 
$\Psi_{\bf 10}\Psi_{\bf 10}H_{\bf 2}$ and 
$\Psi_{\bf 10}\Psi_{\bar{\bf 5}}\tilde H_{\bar{\bf 2}}\Phi_1$ 
should exist. This requires
\begin{equation}
 2\alpha + \rho =0, \qquad 
\alpha+\beta+\xi+\eta-3(a+c)=0.
\end{equation} 
(iv)~The fields $\chi$ and $\bar\chi$ should be massless
at the ${\cal G}$ breaking scale and play the role of the messenger
fields of the supersymmetry breaking which is assumed to occur in the
$S_\alpha$ sector.
These require both the absence of $\Phi_2\chi\bar\chi$ and the existence 
of the coupling $\Phi_2S_\alpha\chi\bar\chi$. These conditions
can be written as
\begin{eqnarray}
&&\delta+\epsilon+\zeta+2(b+d)\not=0,\qquad 
\delta+\epsilon+\zeta-3(b+d)\not=0,
\nonumber \\
&&\delta+\epsilon+\zeta+\theta +2(b+d)=0,\qquad 
\delta+\epsilon+\zeta+\tau-3(b+d)=0.
\label{mess}
\end{eqnarray}
(v)~The neutrino should be massive and the proton should be stable.
This means that $\Psi_{\bar{\bf 5}}\Psi_{\bf 1}H_{\bf 2}$ and 
$\Phi_1\Phi_2\Psi_{\bf 1}^2$
should exist and $\Psi_{\bf 10}\Psi_{\bar{\bf 5}}^2$ and 
$\Psi_{\bf 10}^3\Psi_{\bar{\bf 5}}$ should be
forbidden \cite{w}. These require 
\begin{equation}
\beta+\gamma+\rho =0, \quad 2\gamma=0, \quad
\alpha+ 2\beta\not=0, \quad 3\alpha+ \beta\not= 0.
\end{equation}
(vi)~The gauge invariant bare mass terms of the fields such as 
$\Psi_{\bar{\bf 5}}H$, $H\bar\chi$, $\tilde H\chi$
should be forbidden.\footnote{We cannot forbid the bare mass terms of
the singlet chiral superfields completely based on the discrete 
symmetry $F^\prime$ alone. Although we might need additional symmetry to 
prohibit it, we do not discuss it further here and we only assume that 
they have no bare mass.} 
These conditions are summarized as,
\begin{equation}
\beta+\rho\not=0, \quad \rho+\delta\not=0,\quad  
\xi+\epsilon +2(a+b)\not=0, \quad
\xi+\epsilon -3(a+b)\not=0. 
\label{con2}
\end{equation}

Here we additionally assume 
that both origin of $\mu$ and $B_\mu$ is in the Higgs coupling with
$S_1$ in order to embed our scenario discussed in the previous section 
into this unified model. 
For this realization we introduce a term $\Phi_1S_1H_{\bf 2}\tilde
H_{\bf 2}$ in the superpotential.\footnote{
If we make the Higgs doublets couple to $S_2$ instead 
of $S_1$, $\mu$ seems not to be large enough to satisfy the radiative 
symmetry breaking condition.} The condition for the existence 
of such a term can be written as
\begin{equation}
\rho+\xi+\eta+\theta-3(a+c)=0.
\label{con3}
\end{equation}
Every condition above should be understood up to modulus $n$ when we take
$F^\prime=Z_n$.

We can easily find an example of the consistent solution 
for these constraints (\ref{con1}) -- (\ref{con3}).
In order to show its existence concretely, we give an
example here. If we take $F^\prime=Z_{20}$, these condition can be
satisfied under the charge assignment,\footnote{
We have not taken account of the anomaly of $F^\prime$ here.
Although this anomaly cancellation might require the introduction of
new fields and impose the additional constraints on the charges, 
it does not affect the result of the present study of the model.} 
\begin{eqnarray}
&&\alpha=\eta=1,\quad \rho=\delta=\epsilon=-c=-2, \quad
\xi=\zeta=-\delta=-a=-b=3, \nonumber \\ 
&&\theta=-5, \quad  d=6 ,\quad \beta=-8, 
\quad \gamma=\tau=10. 
\end{eqnarray}
It should be noted that the different singlet fields $S_{1,2}$ are 
generally required for the couplings to $\chi$ and $\bar\chi$, which 
play a role of messengers of the supersymmetry breaking. 
This feature incidentally comes from the introduction of the 
direct product gauge structure motivated to realize the 
doublet-triplet splitting,
which requires the $F^\prime$ charges of $\chi$ and $\bar\chi$ to satisfy
\begin{equation}
\theta-\tau =-5(b+d)\not=0    \qquad ({\rm mod}~n).
\end{equation}
 
We can now consider the physics at the scale after the symmetry breaking 
due to the VEVs in eq.~(\ref{eqb}). The massless degrees of
freedom are composed of the contents of the MSSM and the fields 
$(q, l)$ and $(\bar q, \bar \ell)$ which come from $\chi({\bf 1}, {\bf 5})$
and $\bar\chi(\bar{\bf 5}, {\bf 1})$ and also the singlet fields $S_{1,2}$.
Thus we can expect the successful gauge coupling unification for these 
field contents in the similar way to the MSSM.
Under the imposition of the discrete symmetry $F^\prime$, 
the superpotential for these fields can be written as
\begin{eqnarray}
&&W_2=h_1\Psi_{\bf 10}\Psi_{\bf 10}H_2 
+ {h_2\Phi_1\over M}\Psi_{\bf 10}\Psi_{\bar{\bf 5}} H_{1}
+{h_3 \Phi_1\over M}S_1H_1H_2 \nonumber \\
&&\hspace*{1cm}+h_4\Psi_{\bf 1}\Psi_{\bar{\bf 5}}H_2 
+ {h_5\Phi_1\Phi_2\over M}\Psi_{\bf 1}\Psi_{\bf 1} \nonumber \\
&&\hspace*{1cm}+{\lambda_1\Phi_2 \over M} S_1q\bar q 
+{\lambda_2\Phi_2\over M} S_2\ell\bar\ell, 
\label{eqc}
\end{eqnarray}
where $M$ is the effective unification scale.
We use the usual notation in the MSSM for the doublet Higgs fields 
such as $H_1\equiv \tilde H_{\bar{\bf 2}}$ and $H_2\equiv  H_{\bf 2}$.
The several terms can be suppressed by the additional
factors $\epsilon_{1,2}\equiv\langle\Phi_{1,2}\rangle/M$
coming from the VEVs $\langle\Phi_1\rangle$ and $\langle\Phi_2\rangle$
given in eq.~(\ref{eqb})
since each term is controlled by the discrete symmetry $F^\prime$.
This feature makes several terms phenomenologically favorable.
For example, the second term in the first line which includes the MSSM
relevant terms seems to be favorable to explain the hierarchy between the 
masses of top and bottom quarks for the various values of $\tan\beta$.
The mass hierarchy between the top quark and the bottom quark 
requires that $\epsilon_1$ should be $O(10^{-2})$ or larger.
This feature also causes the favorable effects on the second line 
which is relevant to the neutrino masses. 
In fact, if the VEVs $\langle\Phi_1\rangle$ and $\langle\Phi_2\rangle$
take the suitable values so as to be $\epsilon_1\epsilon_2=O(10^{-2})$, 
the right-handed neutrinos $\Psi_{\bf 1}$
can have the mass of $O(10^{13})$~GeV which is suitable to explain the
experimental data for the solar and atmospheric neutrinos.

In the last line of eq.~(\ref{eqc}), as we expected, 
$q,~\bar q$ and $\ell,~\bar\ell$ 
couple with the different singlet fields $S_{1,2}$. 
Thus the messenger sector assumed in the previous section is realized.
The last term in the first line can be an 
origin of the $\mu$ and $B_\mu$ terms, since both the scalar component 
and $F$-component of $S_1$ are assumed to get the VEVs. 
Both $\mu$ and $B_\mu$ in eq.~(\ref{eqd}) can be induced by taking
$\lambda_\mu=h_3\epsilon_1$.
In fact, if we assume 
\begin{equation}
\epsilon_1= O(10^{-2}), \quad 
\langle S_1\rangle=O(10^5)~{\rm GeV}, \quad 
\langle F_{S_1}\rangle=O(10^9)~{\rm GeV}^2,
\label{eqx}
\end{equation}
we can consistently obtain an suitable values of $B_\mu$ and $\mu$ 
for the radiative electroweak symmetry breaking which has been discussed 
in the previous section. 
To satisfy the neutralino mass bound, however, we need to introduce 
an additional origin for $\mu^\prime$. 
If we can introduce such an origin as $\mu^\prime=O(100)$~GeV, the
radiative symmetry breaking condition is expected to be easily satisfied 
based on the analysis in the previous section.
The new origin may be given by the nonrenormalizable couplings among the 
Higgs chiral superfields and the singlet chiral superfield $N$
whose scalar potential has a negative curvature due to the K\"ahler
potential interaction \cite{mgm1a,mgmrev}. We consider the following 
terms in the effective Lagrangian,
\begin{equation}
\int d^4\theta S_1^\dagger S_1N^\dagger N +\left\{\int d^2\theta 
\left({\Phi_1^p\Phi_2^p\over M^{2p+m}} N^{m+3} +
{\Phi_1^{q+1}\Phi_2^q\over M^{2q+n}} N^n H_1H_2 \right) 
+ {\rm h.c.} \right\},
\label{eff}
\end{equation} 
where each term should be determined by the discrete symmetry presented in
Table~1. From these terms the additional contribution $\mu^\prime$ 
to the $\mu$ term is yielded at the tree-level as 
\begin{equation}
\mu^\prime=\left({1\over (m+2)(m+3)^2}\right)^{n\over 2m+2}
\epsilon_1^{q+1-{pn\over m+1}}\epsilon_2^{q-{pn\over m+1}}
F_{S_1}^{n\over m+1}M^{-2n+m+1\over m+1}.
\end{equation}
If $2n=m+1$ is satisfied \cite{mgm1a,mgmrev}, the magnitude of 
this $\mu^\prime$ is
approximately expressed as $\epsilon_1^{q+1-p/2}\epsilon_2^{q-p/2}
\sqrt{F_{S_1}}$. Therefore, we can obtain $\mu^\prime$ of $O(100)$~GeV by
assuming $p=q=0$ and taking account of eq.~(\ref{eqx}).
Since $B_\mu$ is not generated at the tree-level along with this
$\mu^\prime$, the dominant $B_\mu$ comes from $\Phi_1 S_1H_1H_2$ in
eq.~(\ref{eqc}) and then our result for the radiative symmetry breaking 
obtained in the previous section can be directly applicable to this model.  
It should be also noted that the effective Lagrangian (\ref{eff}) with
$p=q=0$ can be
constructed on the basis of the discrete symmetry $F^\prime$ 
by defining the charge of $N$ as $\omega=-5$ in the case of $n=1$ and $m=1$.   
  
Finally we should comment on the relation to the mass eigenvalues 
and the mixings of quarks and leptons. 
We would like to stress that the existence of the suppression
factor $\epsilon_1$ is favorable for the explanation of the masses
of quarks and leptons as mentioned below eq.~(\ref{eqc}). 
The value of $\epsilon_1$ is constrained by the masses of a bottom 
quark and a $\tau$ lepton.
If we impose $\tan\beta~{^>_\sim}~2$ which is required by the neutral 
Higgs boson mass constraint, $\epsilon_1~{^>_\sim}~10^{-2}$ should 
be satisfied. This is consistent with the condition given in 
eq.~(\ref{eqx}) and also with the neutrino oscillation data. 
If we introduce Frogatt-Nielsen flavor U(1) symmetry into this model along
the line of \cite{sue1}, the qualitatively satisfactory mass eigenvalues 
and mixing angles for quarks and leptons are expected to be derived. 
We will discuss this subject in other place.

\section{Summary}
We have investigated the $\mu$ problem and the radiative symmetry
breaking in the extended GMSB scenario, which can be derived, for example, 
from the supersymmetric unified SU(5) model with the 
doublet-triplet splitting.
The model may be constructed through the deconstruction by extending the
gauge structure into the direct product group 
SU(5)$^\prime\times$SU(5)$^{\prime\prime}$. 
The low energy spectrum is the one of the MSSM with the 
additional chiral superfields which can play a role of messengers 
in the GMSB.
The discrete symmetry forces the color triplet and color singlet
messengers to couple to the different singlet chiral superfields whose 
scalar and auxiliary components are assumed to get the VEVs due to 
the hidden sector dynamics.

In such a model the direct coupling between the doublet Higgs fields and the 
one of these singlet
fields is allowed but suppressed due to this discrete symmetry.
This coupling can give the origin of both $\mu$ and $B_\mu$ terms.
Since the model has two scales which are relevant to the 
supersymmetry breaking and the superpartner masses depend on both of them, 
the induced $\mu$ and $B_\mu$ can be consistent
with the radiative electroweak symmetry breaking.
This aspect is largely different from the ordinary minimal GMSB
scenario and it may present a new solution for the $\mu$ 
problem in the GMSB scenario at least from the viewpoint of the 
radiative symmetry breaking. 
However, to make the model consistent with the experimental
bounds for the masses of superpartners, it seems to be required to
introduce the additional contribution to the $\mu$ term.
Some interesting features different from the ordinary GMSB 
appear in the spectrum of the superpartners.
The mass difference between the colored and color singlet superpartners 
tends to be smaller in comparison with the ordinary GMSB scenario and also
the gaugino masses become non-universal generally.
The next lightest superparticle can be always the neutralino. 
The gauge coupling unification scale may be pushed upwards somewhat.

Further phenomenological study of this kind of model seems to be worthy since
it is constructed on the basis of the reasonable motivation to solve 
the doublet-triplet splitting problem in the grand unified model.

\vspace{.3cm}
\noindent
This work is supported in part by a Grant-in-Aid for Scientific 
Research (C) from Japan Society for Promotion of Science
(No.~14540251) and also by a Grant-in-Aid for Scientific 
Research on Priority Areas (A) from The Ministry of Education, Science,
Sports and Culture (No.~14039205).

%%%%%%%%%%%%%%references%%%%%%%%%%%%%

\end{document}